\documentclass[seceq,supplement]{ptptex}

\usepackage{graphicx}

\title{Supersolid phases of hardcore bosons on the square lattice: 
Correlated hopping, 
next-nearest neighbor hopping and frustration}

\author{
Frederic \textsc{Mila}$^{1,}$\footnote{E-mail: frederic.mila@epfl.ch}, 
Julien \textsc{Dorier}$^{1,}$\footnote{E-mail: julien.dorier@epfl.ch}, 
and Kai Phillip \textsc{Schmidt}$^{2,}$\footnote{E-mail: schmidt@fkt.physik.tu-dortmund.de}
}
\inst{$^1$ Institute of Theoretical Physics, \'{E}cole Polytechnique F\'{e}d\'{e}rale de Lausanne,\\CH-1015 Lausanne, Switzerland\\
      $^2$ Lehrstuhl f\"ur theoretische Physik, Otto-Hahn-Stra\ss e 4,\\D-44221 Dortmund, Germany}

\abst{
We discuss the appearance of supersolid phases for interacting hardcore bosons on the square lattice when, in addition to the standard nearest neighbor hopping and repulsion,  correlated or next-nearest neighbor hopping is present. Having in mind dimer-based quantum magnets in a field described by effective bosonic models of this kind, we put special emphasis on a comparison between the different cases of relative signs of the kinetic processes, which correspond to unfrustrated or frustrated magnetic models. In the unfrustrated case, we compare Quantum Monte Carlo simulations with a mean-field (classical) approach, which is shown to give qualitatively correct results. Using this classical approach for the frustrated case, 
we find that the phase diagram is generically richer than in the unfrustrated case.  We also investigate in detail the differences between standard next-nearest neighbour and correlated hopping over the diagonal, with the conclusion that both cases are similar if checkerboard order is present at half-filling, while a supersolid phase can be stabilized without any adjacent solid phase only in the case of correlated hopping. 
}

\begin{document}

\maketitle

\section{Introduction}
The identification of exotic states of quantum matter in microscopic models is an important issue in
current research on strongly correlated quantum systems. The existence of a lattice supersolid (SS) which
simultaneously displays crystalline order (solid) and long-range phase coherence
(superfluid, SF) has been definitely established recently thanks to extensive Quantum Monte Carlo (QMC) simulations
of a hard-core boson model with nearest neighbor repulsion on the triangular lattice in the
context of cold atoms loaded into optical lattices.~\cite{wesse05,heida05,melko05} 

The possibility to realize a supersolid phase in dimer-based quantum magnets, first pointed out by Momoi and Totsuka in the context of SrCu$_2$(BO$_3$)$_2$~\cite{momoi00}, has been further investigated lately~\cite{ng06,sengu07,laflo07}. It relies on the description of a polarized triplet on a dimer as a hard-core boson, a convenient language we use throughout this paper. 

A direct consequence of frustration in effective models for dimer-based quantum magnets in a magnetic field is the supression of the otherwise dominant nearest neighbor hopping. In this situation, correlated hopping terms can become the most important kinetic processes. It has been shown recently by QMC simulations combined with a semi-classical approach (SCA) that an unfrustrated correlated hopping leads to large supersolid phases\cite{schmi08}. At the same time, it has been shown by QMC that standard longer-range hoppings give similar supersolid phases once the corresponding solid phase is stabilized\cite{chen08}. For realistic systems, the kinetic couplings are usually frustrated which prevents the use of QMC. In this work, we therefore want to further explore the influence of frustration and of the nature of the kinetic processes on the appearance of kinetically-driven supersolid phases.   

\section{Model}

\begin{figure}[t!]
  \begin{center}
    \includegraphics{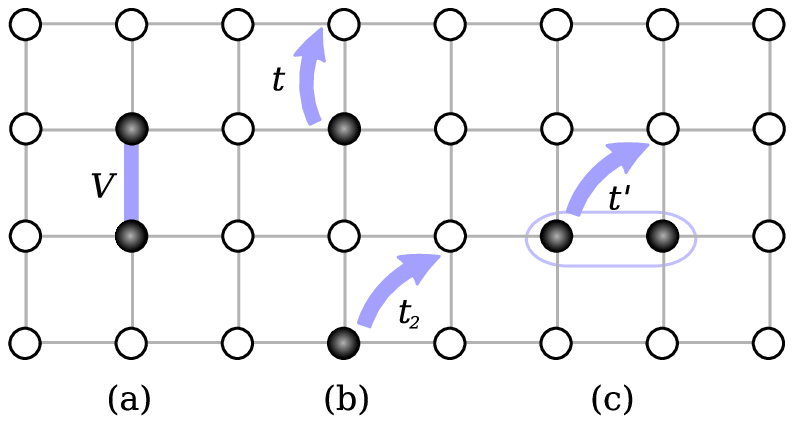}
  \end{center}
  \caption{\label{fig:srcu2bo32:simplemodel:hamiltonian} Illustration of the
  processes of the Hamiltonian Eq.~\ref{Hamiltonian}: 
  nearest neighbor interaction $V$ (a), nearest neighbor and next-nearest neighbor hopping $t$ and $t_2$ (b) and 
  correlated hopping $t'$ (c). The particles are depicted as black circles, while
  white circles correspond to empty sites. }
\end{figure}

We study hard-core bosons on the square lattice defined by the Hamiltonian:
\begin{eqnarray}
 H_{\mu,t,t_2,t',V}&=&-\mu\sum_i n_i-t\sum_{\langle i,j\rangle} \left( b^\dagger_i b^{\phantom{\dagger}}_j+{\text{h.c.}}\right)+V\sum_{\langle i,j\rangle} n_i n_j \nonumber\\
 &&-t_2\sum_{\langle\langle i,j\rangle\rangle} \left( b^\dagger_i b^{\phantom{\dagger}}_j+{\text{h.c.}}\right) -t^\prime\sum_i\sum_{\delta=\pm x;\delta^\prime=\pm y} n_i\left[b^\dagger_{i+\delta}b^{\phantom{\dagger}}_{i+\delta^\prime}+{\it h.c.}\right] 
\label{Hamiltonian}
\end{eqnarray}
where $n_i=b^\dagger_i b^{\phantom{\dagger}}_i\in\{0,1\}$ is the boson density at site $i$, $\mu$ the chemical potential, $t$ the nearest neighbor hopping amplitude, $t_2$ the next-nearest neighbor hopping amplitude, $t^\prime$ the amplitude of the correlated hopping, and $V$ a nearest neighbor repulsion. All the couplings are illustrated in Fig.~\ref{fig:srcu2bo32:simplemodel:hamiltonian}.
 
The amplitude $t_2$ describes direct hopping over the diagonal of the square lattice. In contrast, the correlated hopping term describes a process where a particle can hop along the diagonal of a square plaquette provided there is a particle on one of the other two sites of the plaquette. In this paper we restrict the discussion to the cases of pure correlated hopping $(t_2=0;t'\neq 0)$ and pure direct hopping $(t_2\neq 0;t'=0)$. 

The case $(t_2=0;t'=0)$ has already been thoroughly investigated~\cite{batro00,schmid02}. For strong enough repulsion, an insulating phase with checkerboard (CB) order appears at half-filling. The phase diagram is symmetric about $n=1/2$ in that case due to particle-hole symmetry, and the transition from the solid to the superfluid phase is first order with a jump in the density~\cite{batro00}.

 The Hamiltonian Eq.~\ref{Hamiltonian} is invariant under the translation of one lattice spacing (in $x$ and $y$ direction) as well as under U(1) gauge transformations.  We note that by changing all operators $b^{(\dagger )}$ to $-b^{(\dagger )}$ on one sublattice, the Hamiltonian $H_{\mu,t,t_2,t',V}$ is transformed to $H_{\mu,-t,t_2,t',V}$. It is therefore sufficient to consider the case $t>0$. However, the relative sign between the nearest neighbor hopping $t$ and the kinetic processes over the diagonal $(t_2;t')$ is important. If $t$ and $(t_2;t')$ are positive, both couplings are ferromagnetic in the magnetic language (see below for details) and there is no frustration. In contrast, $(t_2;t')<0$ represents an antiferromagnetic coupling and the different kinetic processes are frustrated. Additionally, the full Hamiltonian is not particle-hole symmetric. Only in the case of a vanishing correlated hopping particles and holes behave in an equivalent manner. 

The unfrustrated case (all kinetic couplings positive) is accessible by QMC simulations. Consequently, the two cases $(t_2=0;t'\neq 0)$\cite{schmi08} and $(t_2\neq 0;t'=0)$\cite{chen08} have been studied recently. In both cases, a large supersolid phase is detected once the checkerboard solid is present. But the two kinetic processes behave differently in the absence of the solid phase which will be further illustrated below. 

\section{Method}

\begin{figure}[t!]
  \begin{center}
    \includegraphics[width=0.9\textwidth]{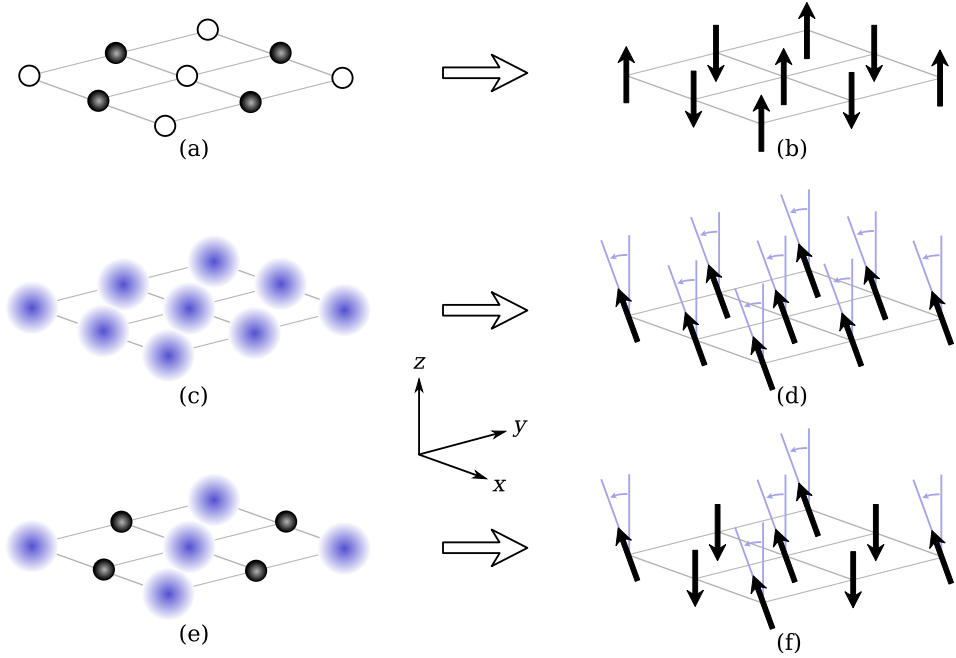}
  \end{center}
  \caption{\label{fig:srcu2bo32:exactdiagonalization_supersolids_sketch_2} Translation of
  bosonic phases into spin language. The checkerboard
  solid (a) becomes a N\'eel state (b). A superfluid (c) becomes a
  ferromagnetic state with a non-zero projection onto the $xy$-plane (d). A
  supersolid (e) becomes a state with N\'eel order but with the spins pointing
  up having a non-zero projection onto the $xy$-plane (f). In this
  sketch, black circles denote completely filled sites, blurred circles
  denote partially filled sites with a non-zero superfluid order parameter
  and black arrows denote expectation values of the spin operators.}
\end{figure}

In order to map out the whole phase diagram, our approach consists in using a classical approach (CA). 
This approximation can be done by first mapping
the hard-core boson operators onto spins 1/2 operators with the
Matsubara-Matsuda transformation {\cite{matsubara56}}:
\begin{eqnarray}
  n_r & = & \frac{1}{2} - S_r^z  \label{equ:srcu2bo32:simplemodel:0050}\\
  b_r & = & S_r^+  \label{equ:srcu2bo32:simplemodel:0060}\\
  b^{\dag}_r & = & S_r^-  \label{equ:srcu2bo32:simplemodel:0070}
\end{eqnarray}
where $S_r^{\pm} = S_r^x \pm S_r^y$ and $n_r = b_r^{\dag} b_r$. After this
transformation, which is exact, Hamiltonian Eq.~\ref{Hamiltonian} becomes
\begin{eqnarray}
  H & = & \frac{1}{2} N (V - \mu) + (\mu - 2 V) \sum_r S^z_r \nonumber\\
  &  & + \sum_r \sum_{\delta = \pm x, \pm y} \left( - t \left( S^x_r S^x_{r +
  \delta} + S^y_r S^y_{r + \delta} \right) + \frac{V}{2} S^z_r S^z_{r +
  \delta} \right) \nonumber\\
  &  & - (t_2+t') \sum_r \sum_{\delta = \pm x} \sum_{\delta' = \pm y} \left( S^x_{r
  + \delta} S^x_{r + \delta'} + S^y_{r + \delta} S^y_{r + \delta'} \right)
  \nonumber\\
  &  & + 2 t' \sum_r \sum_{\delta = \pm x} \sum_{\delta' = \pm y} S^z_r
  \left( S^x_{r + \delta} S^x_{r + \delta'} + S^y_{r + \delta} S^y_{r +
  \delta'} \right) \quad . \label{equ:srcu2bo32:simplemodel:0080}
\end{eqnarray}
For $(t_2=0;t' = 0)$, one obtains a Heisenberg XXZ model in a magnetic field. 
The diagonal and correlated hoppings give a next-nearest neighbor XX-type interaction while
 the correlated hopping leads additionally to a three-spin term. With this transformation, an
empty site is replaced by a spin up and an occupied site is replaced by a spin
down. Therefore the density $n$ is replaced by the magnetization per site $m =
1/2 - n$ and the chemical potential $\mu$ is replaced by a magnetic field $B
= \mu - 2 V$. The solid order parameter becomes
\begin{equation}
  S (k \neq 0) = \frac{1}{N^2} \sum_{r, r'} \langle n_{r'} n_r \rangle e^{i k
  (r' - r)} = \frac{1}{N^2} \sum_{r, r'} \langle S_{r'}^z S_r^z \rangle e^{i k
  (r' - r)} \quad .
\end{equation}
Since the CA and the SCA explicitly break the U(1) gauge symmetry (rotation of all the spins around the $z$ axis ) except for insulating
phases, $\langle b_r \rangle$ can be used as a superfluid order parameter. With the
Matsubara-Matsuda transformation it becomes $\langle b_r \rangle = \langle
S_r^+ \rangle$. Its modulus is given by the length of the spin projection onto
the $xy$-plane
\begin{equation}
  | \langle b_r \rangle | = \sqrt{\langle S_r^x \rangle^2 + \langle S_r^y
  \rangle^2}
\end{equation}
while its phase is given by the angle between the $x$-axis and the spin
projection onto the $xy$-plane. The bosonic phases can be easily translated
into spin language. As shown in Fig.~\ref{fig:srcu2bo32:exactdiagonalization_supersolids_sketch_2}, an insulating
state with checkerboard order (a) becomes a N\'eel state (b), a superfluid
state (c) becomes a ferromagnetic state with all the spins having a non-zero
projection onto the $xy$-plane (d). Finally, a supersolid phase with
checkerboard order (e) becomes a state with a N\'eel order but with all the
spins pointing up which have a non-zero projection onto the $xy$-plane (f).
Note that depending on the density, the spins pointing down in the supersolid
could also have non-zero $xy$-plane components. 

The CA consists in treating the spin operators ${\bf S}_r$
as classical vectors of length $1/2$ {\cite{scalettar95}}. Within this
approximation, the ground state is given by the arrangement of spins $\{
{\bf S}_r^{\text{class}} \}$ which minimizes the classical Hamiltonian. The minimization is then performed
numerically on a four-site square cluster with periodic boundary conditions which is sufficient for the model under study to
allow for broken translational symmetry.

A possible extension of the method consists in evaluating the quantum fluctuations around
the classical ground state with a standard linear spin-wave approach.
The linear spin-wave approximation is based on the standard Holstein-Primakoff transformation of spins into bosons. It consists in replacing the square
root by its lowest order expansion, an approximation which is justified if  the quantum
fluctuations around the classical ground state are small enough, in which case
$\langle a_r^{\dag} a_r \rangle \ll 2 S$ in the ground state. The next step 
of the linear spin-wave
approximation is to neglect all the terms in the Hamiltonian which contains
more than two bosonic operators. The resulting Hamiltonian is quadratic in
bosonic operators and can be diagonalized by using a Fourier transformation
and a Bogoliubov transformation. It is then possible to evaluate the ground
state expectation value of any observable, and in particular the order
parameters, the magnetization per site and the energy.

If the classical ground state is unique, or if its degeneracy is due to a
symmetry of the Hamiltonian, this procedure is in principle sufficient.
However, if the classical ground state is degenerate and the degeneracy is not
related to a symmetry, one can expect the quantum fluctuation to lift this
degeneracy. One could also expect the quantum fluctuations to change the energy of a higher energy
classical state in such a way that it could become a ground state. It is sufficient to check 
whether the
classical states which are stationary points of the classical energy can give a
semi-classical ground state. Therefore in principle, the semi-classical
treatment should be performed for every state which is a stationary point of
the classical energy, the resulting semi-classical energies should be
compared, and the minimum should be extracted. In practice, a numerical method
is used to find several ($\sim 100$) different classical states which are
local minima of the classical energy. For each of these states the
semi-classical treatment is performed and only the results which minimize the
semi-classical energy are kept as ground states.

In this work we have used the SCA only for the case $(t>0;t'>0)$. In that case, we have compared the CA and SCA approximations, with the result that only the first order transitions in the phase diagram can be slightly shifted, while second order transitions do not change. 
Since we are mostly interested in the qualitative phase diagram of our model,  we have 
restricted the investigation of the other cases to the CA. Clearly, if one is aiming
at a more quantitative understanding of the different order parameters, a SCA calculation can be very useful.

\section{Effects of correlated hopping}

In the following we study the case $(t_2=0;t'\neq 0)$. We will first discuss the unfrustrated case $t'>0$ which has already been analyzed before by QMC and SCA\cite{schmi08}. This will allow on one hand to discuss the validity of the CA and on the other hand to compare this case to the frustrated one and to the case of normal diagonal hopping.

\subsection{Unfrustrated case $(t>0;t'>0)$}

\begin{figure}[t!]
  \begin{center}
    \includegraphics[width=0.9\textwidth]{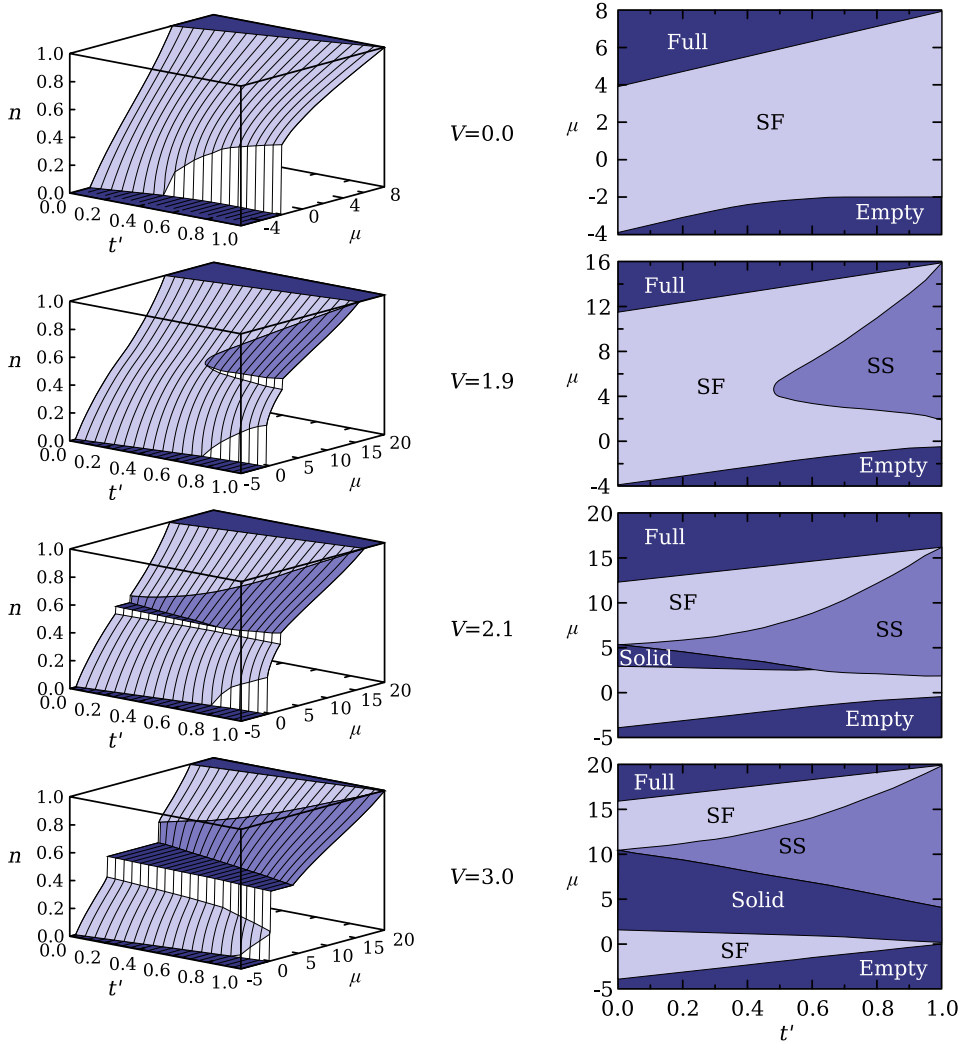}
  \end{center}
  \caption{\label{fig:srcu2bo32:exactdiagonalization_supersolids_phase_diagram_fixed_V}Semi-classical
  results obtained on a $N = 150 \times 150$ lattice. Each row corresponds to
  a given nearest neighbor repulsion $V$. The first column presents the
  density $n$ as a function of the chemical potential $\mu$ and the correlated
  hopping $t'$ at fixed $V$. The second column presents the $\mu - t'$ phase
  diagrams at fixed $V$. White regions denote phase separation, darker
  regions correspond to superfluid phases, supersolid phases and solid
  (insulating) phases, as shown in the second column.}
\end{figure}

\begin{figure}[t!]
  \begin{center}
    \includegraphics[width=0.9\textwidth]{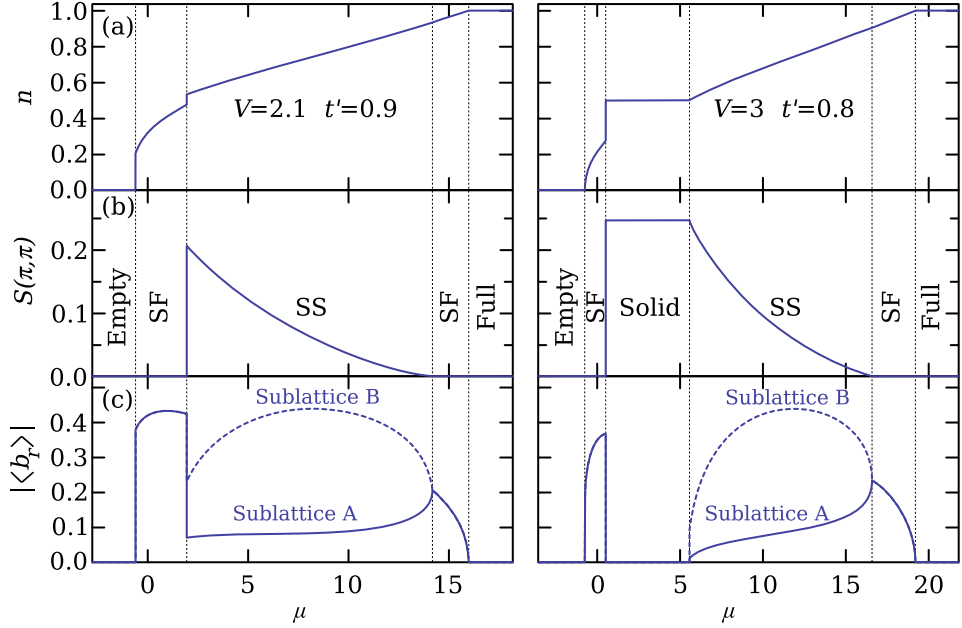}
  \end{center}
  \caption{\label{fig:srcu2bo32:exactdiagonalization_supersolids_orderparameters}
Semiclassical results for (a)
  the density $n$, (b) the static structure factor $S (\pi, \pi)$ and (c) the superfluid
  order parameter $| \langle b_r \rangle |$  for $r$ on the $A$ sublattice
  (solid lines) and $B$ sublattice (dashed lines) as a function of the
  chemical potential for two representative cases. Left panel: supersolid without neighboring solid
  at $t' = 0.9$ and $V = 2.1$. Right panel: $t' = 0.8$ and $V = 3$. }
\end{figure}

The semi-classical results are shown in Fig.~\ref{fig:srcu2bo32:exactdiagonalization_supersolids_phase_diagram_fixed_V}.
The first column presents the density $n$ as a function of the chemical
potential $\mu$ and the correlated hopping $t'$ at fixed nearest neighbor
repulsion $V$. The corresponding $\mu - t'$ phase diagrams at fixed nearest
neighbor repulsion $V$ are shown in the second column. Each row corresponds to
a different nearest neighbor repulsion. When $V = 0$ (first line of Fig.~\ref{fig:srcu2bo32:exactdiagonalization_supersolids_phase_diagram_fixed_V}),
the phase diagram is dominated by a superfluid phase. There is a phase
separation region at low $n$ and large $t'$, in agreement with exact
diagonalization and QMC results\cite{schmi06}. The superfluid
phase corresponds to a ferromagnetic state, with a non-zero component in the
$xy$-plane, as depicted in Fig.~\ref{fig:srcu2bo32:exactdiagonalization_supersolids_sketch_2} (c) and (d). For
$0 < V < 2$, a supersolid phase appears in the large $\mu$ and large $t'$
region of the phase diagram and extends towards lower $t'$ when $V$ increases,
as shown in the second line of Fig.~\ref{fig:srcu2bo32:exactdiagonalization_supersolids_phase_diagram_fixed_V}. In
the supersolid phase, the state is close to a N\'eel state, but the spins
acquire a ferromagnetically ordered non-zero component in the $xy$-plane.
This state is similar to the state depicted in Fig.~\ref{fig:srcu2bo32:exactdiagonalization_supersolids_sketch_2} (e) and (f),
except that the spins pointing down have also a non-zero $x y$ plane
component. For $V \geqslant 2$, as shown in the bottom of Fig.~\ref{fig:srcu2bo32:exactdiagonalization_supersolids_phase_diagram_fixed_V}, a
$n = 1 / 2$ insulating (solid) phase appears from $t' = 0$ when $V = 2$ and
extends towards larger $t'$ when $V$ increases. In this insulating phase, which
appears as a plateau in the $n$ versus $\mu$ curves, the system is in the
N\'eel state shown in Fig.~\ref{fig:srcu2bo32:exactdiagonalization_supersolids_sketch_2} (a) and (b). At
$V = 2$, the supersolid region reaches up to the limiting case $t' = 0$, and becomes larger
when $V$ increases. The order parameters are shown in Fig.~\ref{fig:srcu2bo32:exactdiagonalization_supersolids_orderparameters} for two
reprentative cases $t' = 0.9$ $V = 2.1$ (left panel) and $t' = 0.8$ $V = 3$
(right panel).

\begin{figure}[t!]
  \begin{center}
    \includegraphics[width=0.9\textwidth]{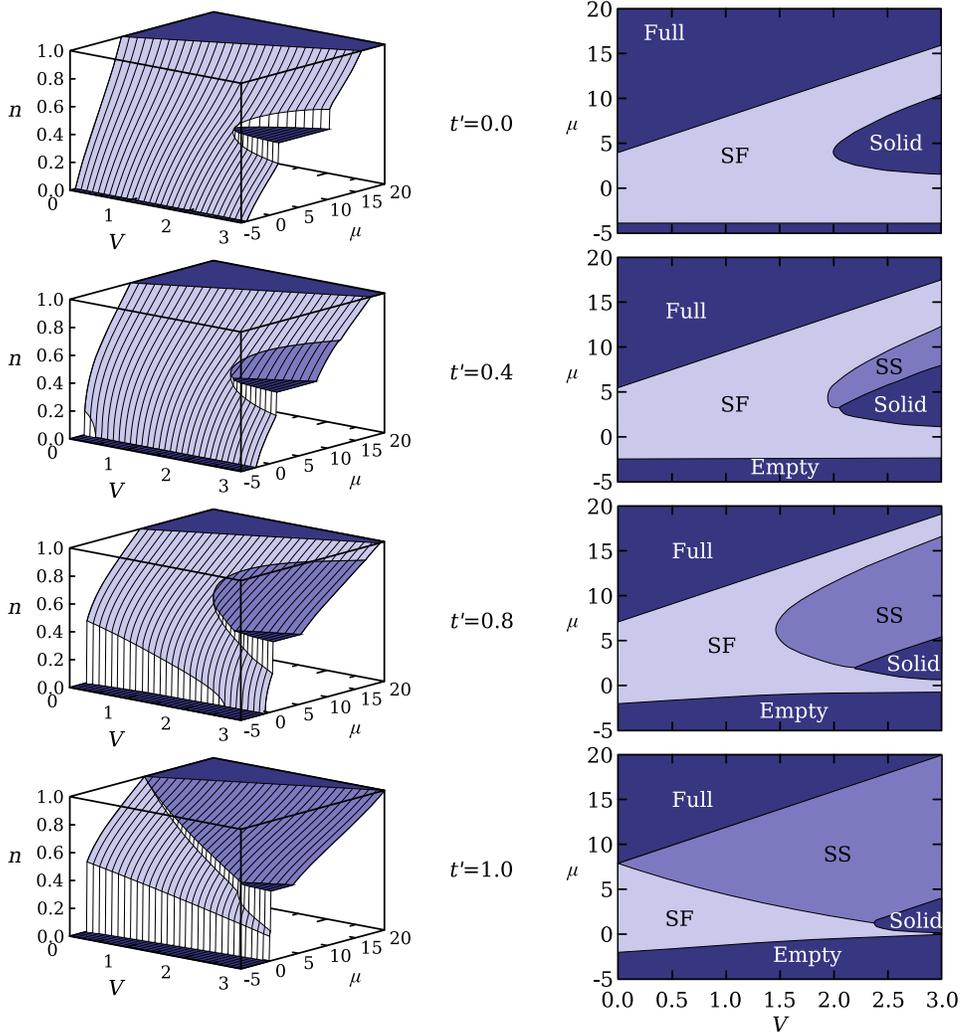}
  \end{center}
  \caption{\label{fig:srcu2bo32:exactdiagonalization_supersolids_phase_diagram_fixed_tp}Semi-classical
  results obtained on a $N = 150 \times 150$ lattice. Each row corresponds to
  a given nearest correlated hopping $t'$. The first column presents the
  density $n$ as a function of the chemical potential $\mu$ and the nearest
  neighbor repulsion $V$ at fixed $t'$. The second column shows the $\mu -
  V$ phase diagrams at fixed $t'$. White regions denote phase separation,
  darker regions correspond to superfluid phases, supersolid phases and solid
  (insulating) phases, as shown in the second column.}
\end{figure}

Fig.~\ref{fig:srcu2bo32:exactdiagonalization_supersolids_phase_diagram_fixed_tp}
presents essentially the same information as Fig.~\ref{fig:srcu2bo32:exactdiagonalization_supersolids_phase_diagram_fixed_V},
however it is useful in order to better understand the whole phase diagram. It
displays the same quantities as Fig.~\ref{fig:srcu2bo32:exactdiagonalization_supersolids_phase_diagram_fixed_V} but
as a function of the nearest neighbor repulsion $V$ and at fixed correlated
hopping $t'$.

These phase diagrams show two interesting features. Firstly, the large $V$
and large $t'$ region is dominated by a very large supersolid phase that
appears for densities above 1/2. Secondly, this supersolid phase is stable
even at very small $V$, where the checkerboard solid is not stabilized. This
feature is strongest at $t' = 1$, where the SCA
predicts a stable supersolid phase for all $V > 0$ while the solid phase is
stabilized only when $V$ is larger than a critical value $V_c^{S S} \simeq
2.3$ (see Fig.~\ref{fig:srcu2bo32:exactdiagonalization_supersolids_phase_diagram_fixed_tp}).
In other words, the correlated hopping can stabilize large supersolid phases.
But this effect is so strong that for a given parameter set ($t', V$), the
correlated hopping can stabilize a supersolid phase even when there is no
neighboring solid phase as $\mu$ is changed. Note that correlated hopping
appears to be crucial for this physics to be realized\cite{schmi08}. Indeed, the same model with a standard 
next-nearest neighbor hopping never stabilizes a supersolid without an adjacent solid phase (see below).

Upon increasing the chemical potential $\mu$, the SCA
predicts first order transitions from superfluid to solid and from superfluid
to supersolid, while the transitions from supersolid to superfluid and solid
to supersolid are second order. Note that the second order transition from
solid to superfluid for $V > 2$ and $t' > 0$ becomes a first order transition
from solid to superfluid when $t' = 0$ (see Fig.~\ref{fig:srcu2bo32:exactdiagonalization_supersolids_phase_diagram_fixed_V}).


\subsubsection{Comparison CA and SCA}

\begin{figure}[t!]
  \begin{center}
    \includegraphics{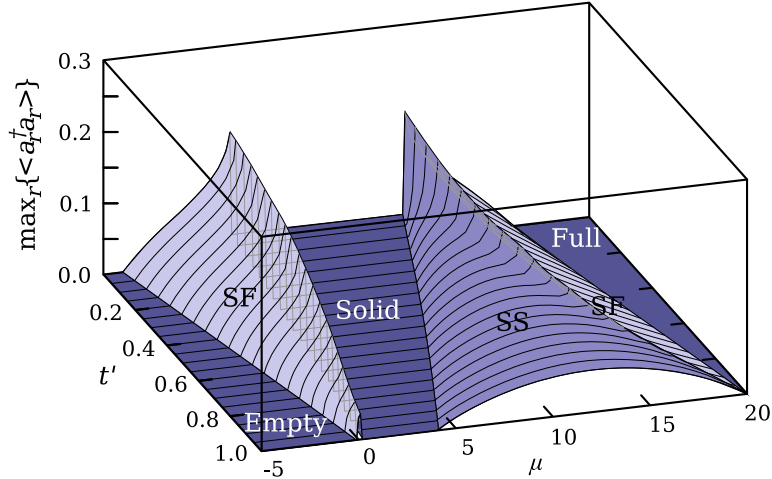}
  \end{center}
  \caption{\label{fig:srcu2bo32:exactdiagonalization_supersolids_quantum_fluctuations}Maximum
  number of Holstein-Primakoff bosons $\max_r \{\langle a_r^{\dag} a_r
  \rangle\}$ obtained with the SCA on a $N = 150
  \times 150$ lattice as a function of the chemical potential $\mu$ and the
  correlated hopping $t'$ for $V = 3.0$. The color coding indicates which
  phase is realized for the corresponding parameters $V, t'$ and $\mu$.}
\end{figure}

The main approximation in the semi-classical treatment of the model consists
in replacing the square root appearing in the Holstein-Primakoff
transformation by its lowest order expansion around $a_r^{\dag} a_r / (2
S) = 0$
\begin{equation}
  \sqrt{1 - \frac{a_r^{\dag} a_r}{2 S}} \simeq 1
\end{equation}
As a thumb rule, this approximation is expected to be qualitatively correct
provided the ground state expectation value $\langle a_r^{\dag} a_r
\rangle$ is small compared to $2 S$. This expectation value has been evaluated
with the other quantities and it satisfies $\langle a_r^{\dag} a_r \rangle <
0.2$ for any $V, t', \mu$ and $r$ which is small compared to $2 S$, even for
spins 1/2. Interestingly, $\langle a_r^{\dag} a_r \rangle$ is larger around
the phase transitions between two phases with and without solid order, as
shown in Fig.~\ref{fig:srcu2bo32:exactdiagonalization_supersolids_quantum_fluctuations}
which presents $\max_r \{\langle a_r^{\dag} a_r \rangle\}$ as a function of
$\mu$ and $t'$ for $V = 3.0$. Therefore, one may expect that the phase
transitions predicted by the SCA are slightly
displaced with respect to the transitions obtained with an exact treatment of
the model.

In order to determine the phase diagram for more complicated models, it is in practice often necessary to restrict oneself to the CA instead of the SCA. It is therefore interesting to know what
differences should be expected. Close to a phase transition, within the
CA, upon increasing the chemical potential $\mu$ the
energy of the first excited state crosses the ground state energy at a
critical value $\mu_c$ and becomes the new ground state. If the transition is
continuous, the two states have to be identical at the transition point and
the effect of the quantum fluctuations will be the same for both states. In
particular their energy will change by the same amount and the transition will
occur at the same $\mu_c$ as in the SCA. By contrast, at a first order transition, both states are different.
Therefore the two energies will in general not change by the same amount and
the intersection will occur at another $\mu_c$ in the SCA. In addition to this effect, one should also expect renormalizations of
all the measured quantities due to the quantum fluctuations.

In conclusion, as long as we are only interested in phase diagrams, the
CA will give the same results as the SCA, except at first order transitions where one should expect small
shifts of the boundaries.

\subsubsection{Comparison SCA and QMC}

\begin{figure}[t!]
  \begin{center}
    \includegraphics[width=0.48\textwidth]{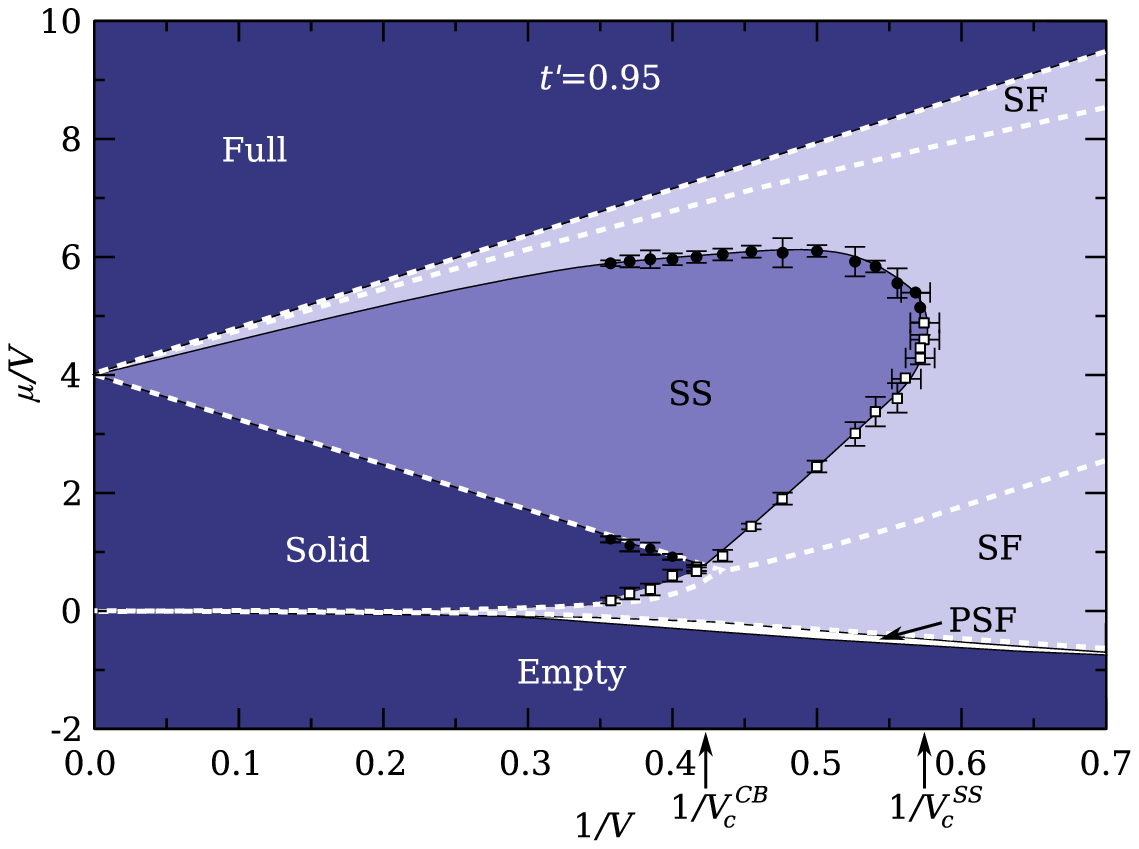}
    \includegraphics[width=0.48\textwidth]{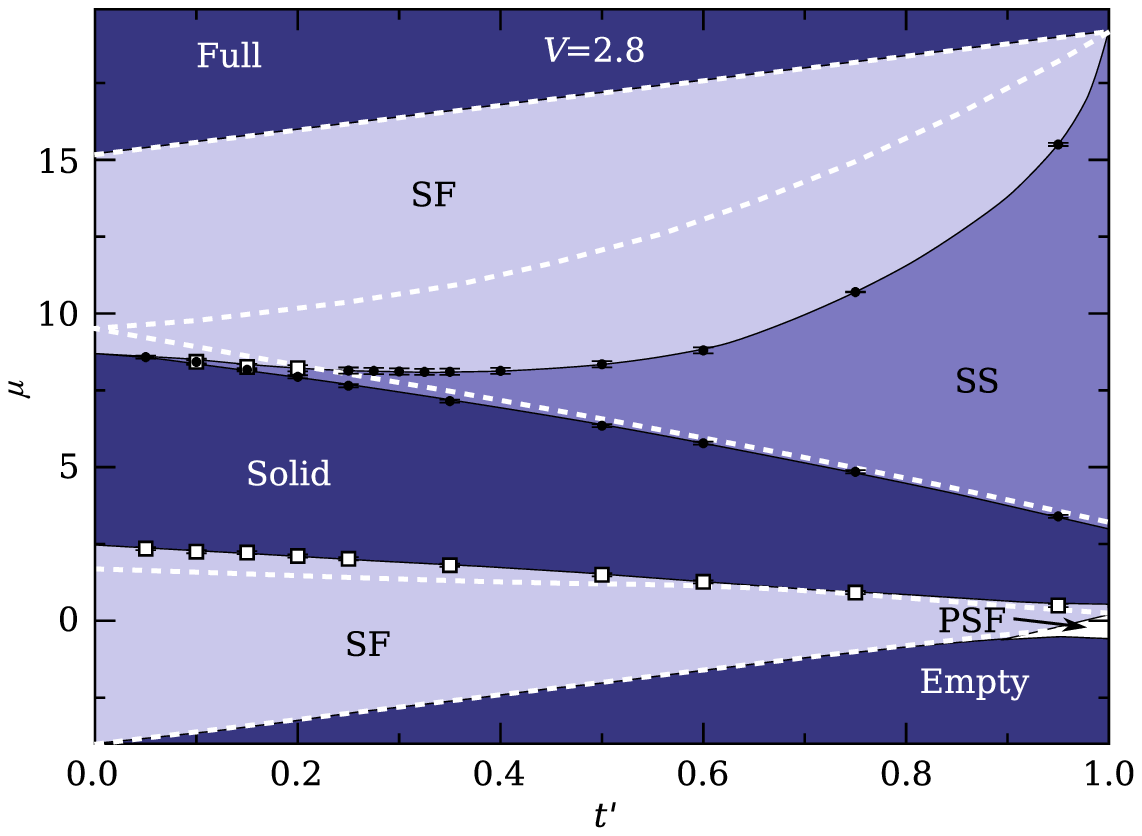}
  \end{center}
  \caption{\label{fig:srcu2bo32:exactdiagonalization_supersolids_qmc_1}{\it Left:} Zero-temperature
  phase diagram for $t' = 0.95$ as a function of $1 / V$ and $\mu / V$. White
  squares (black circles) denote first (second) order phase transitions
  deduced from quantum Monte Carlo. White dashed lines are semi-classical
  results. The other lines are only guide to the eye. {\it Right:} Zero-temperature
  phase diagram for $V = 2.8$ as a function of $t'$ and $\mu$. White squares
  (black circles) denote first (second) order phase transitions deduced from
  quantum Monte Carlo. White dashed lines are semi-classical results. The
  other lines are only guide to the eye. Note that both phase diagrams are taken from Ref.~\citen{schmi08}.}
\end{figure}

Here we would like to compare the SCA results discussed above with unbiased QMC simulations. Two representative phase diagrams 
are shown in Fig.~\ref{fig:srcu2bo32:exactdiagonalization_supersolids_qmc_1}. The left figure shows a phase diagram for fixed correlated 
hopping $t'=0.95$ as a function of $1/V$ and $\mu /V$. It can be clearly seen that the checkerboard solids in SCA and QMC agree almost quantitatively. The SCA only slightly overestimates the solid which is expected since quantum fluctuations are not fully treated in the SCA. 

The second interesting feature emerging out of the SCA calculation is that a supersolid phase can be stabilized by correlated hopping without any adjacent solid phase. In the SCA, this feature is present for any finite value of $V$ at large values of the correlated hopping $t'$. Qualitatively, this scenario is also present in the QMC simulations. There can be indeed a supersolid phase without any adjacent solid phase\cite{schmi08}. But this feature is restricted to a much narrower region in parameter space. One needs for example a finite value of $V=1.74$ to see this effect. The SCA therefore strongly overestimates the supersolid phase in this case.

In the right panel of Fig.~\ref{fig:srcu2bo32:exactdiagonalization_supersolids_qmc_1}, we show the phase diagram for $V = 2.8$ as a function of $t'$ and $\mu$. For these parameters the checkerboard solid is stabilized at half filling and one finds a large supersolid phase upon doping particles. Here all global features in SCA and in QMC are similar up to small overestimates by the SCA.  

\subsection{Frustrated Case $(t>0;t'<0)$}

In the previous section, the phase diagram of Hamiltonian Eq.~\ref{Hamiltonian} 
has been thoroughly investigated for the unfrustrated case ($t>0;t'>0$). In the magnetic language
both kinetic terms act as ferromagnetic couplings in the $xy$-plane. In any superfluid or
supersolid phase, the system can easily minimize the kinetic energy by
ordering the $xy$-components of the spins ferromagnetically. In terms
of bosons, it corresponds to a phase with all the particles having the same
phase.

In contrast, when the kinetic processes are frustrated $(t>0;t'<0)$, the coupling in the $xy$-plane 
is still ferromagnetic for nearest neighbor spins, but becomes
antiferromagnetic for next-nearest neighbor spins. Although in the limits $t
\gg t'$ and $t \ll t'$ the system can still find an ordering which minimizes
the kinetic energy, this is not true anymore when $t \simeq t'$. As a
consequence, one should expect the phases with non-zero $xy$-plane components
of the spins (superfluid and supersolid) to have higher energies than in the
 unfrustrated case, while the energy of the solid phase is almost unchanged. This results
generically in a phase diagram with less superfluid and supersolid regions.

\begin{figure}[t!]
  \begin{center}
    \includegraphics[width=0.9\textwidth]{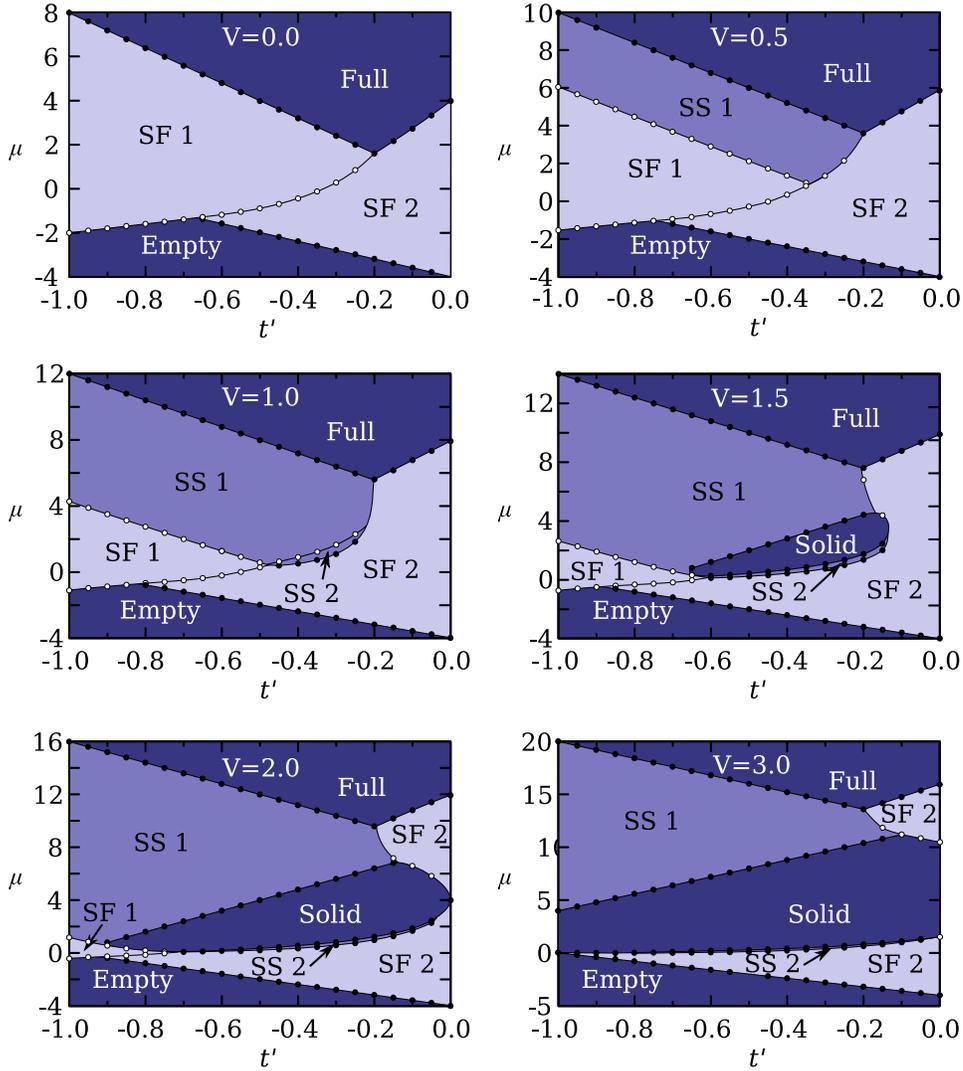}
  \end{center}
  \caption{\label{fig:srcu2bo32:exactdiagonalization_negative_tp_phasediagram}Zero-temperature
  phase diagrams as a function of the correlated hopping $t'$ and the chemical
  potential $\mu$ for various nearest neighbor repulsion $V$. White (black)
  circles corresponds to first (second) order transitions obtained with the
  CA. The lines are only guides to the eye. }
\end{figure}

\begin{figure}[t!]
  \begin{center}
    \includegraphics[width=0.9\textwidth]{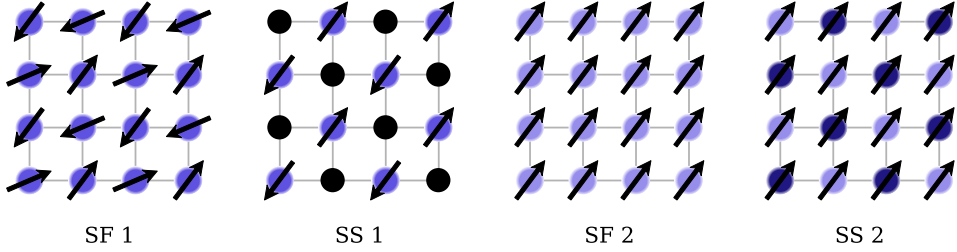}
  \end{center}
  \caption{\label{fig:srcu2bo32:exactdiagonalization_negative_tp_phases}Structure
  of the various superfluid (SF) and supersolid (SS) phases. The density $n_r$
  at a given site $r$ is represented by the intensity of circle, which goes
  from white for $n_r = 0$ to black for $n_r = 1$. When $n_r \neq \{0, 1\}$, the
  particles acquire a superfluid component whose phase is represented by the
  angle between the black arrow and the $x$ axis. In terms of spins, the
  arrows correspond to the $xy$-plane components, while the color of the
  circle corresponds to the $z$ component.}
\end{figure}

\begin{figure}[t!]
  \begin{center}
    \includegraphics[width=0.9\textwidth]{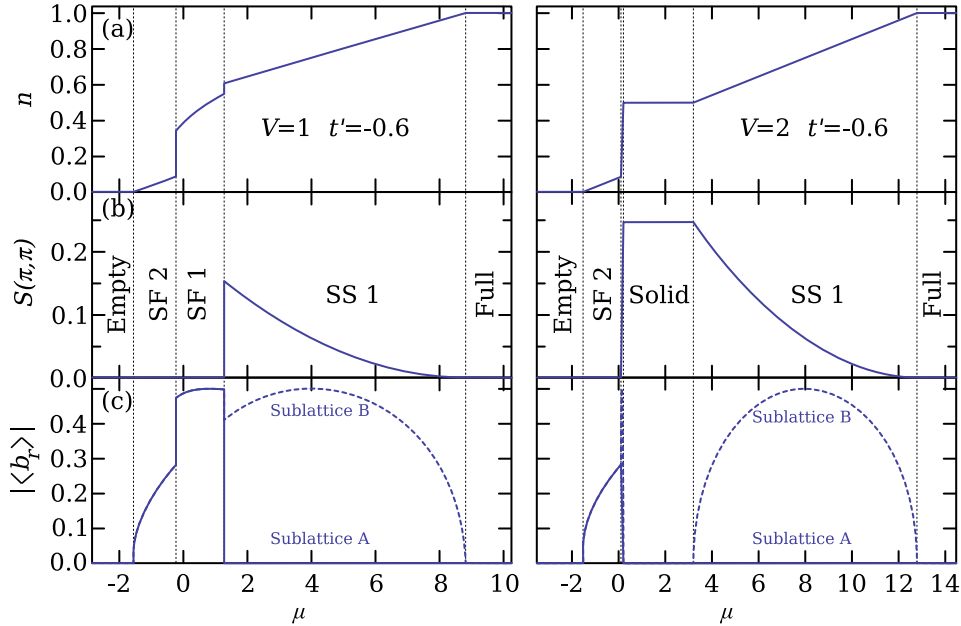}
  \end{center}
  \caption{\label{fig:srcu2bo32:exactdiagonalization_negative_tp_orderparameters}(a)
  Semiclassical results for (a)
  the density $n$, (b) the static structure factor $S (\pi, \pi)$ and (c) the superfluid
  order parameter $| \langle b_r \rangle |$  for $r$ on the $A$ sublattice
  (solid lines) and $B$ sublattice (dashed lines) as a function of the
  chemical potential for two representative cases. Left panel: supersolid without neighboring solid at
  $t' = - 0.6$ and $V = 1$. Right panel: $t' = - 0.6$ and $V = 2$. }
\end{figure}

In order to check these predictions, a CA has been used to
build the zero-temperature phase diagram of Hamiltonian Eq.~\ref{Hamiltonian} 
for $(t>0;t'<0)$. All the
calculations were performed on four-site clusters, but because the frustration
of the model could lead to larger periodicities in the superfluid and
supersolid phases, test calculations with clusters up to eight sites have been
done in all different phases. Similarly to the previous case, the energy scale
is fixed by $t + |t' | = 1$. The resulting phase diagrams as a function of the
correlated hopping $t'$ and the chemical potential $\mu$ for various values of
the nearest neighbor repulsion $V$ are shown in Fig.~\ref{fig:srcu2bo32:exactdiagonalization_negative_tp_phasediagram}, while the
structure of the various superfluid and supersolid phases are sketched in
Fig.~\ref{fig:srcu2bo32:exactdiagonalization_negative_tp_phases}. The
density and the order parameters are shown in Fig.~\ref{fig:srcu2bo32:exactdiagonalization_negative_tp_orderparameters} as a
function of the chemical potential for two reprentative cases $(t'=-0.6;V=1)$ in the left panel and $(t'=-0.6;V = 2)$ in the right panel.

When $V = 0$, the phase diagram is dominated by superfluid phases. However,
because of the competition between hopping and correlated hopping, the phase
diagram is already more interesting than in the unfrustrated case ($t, t' >
0$). In the limit where simple hopping dominates ($t' \simeq 0$), the system
stabilizes a superfluid phase (SF 2) with a ferromagnetic ordering of the $xy$-plane 
components of the spins (i.e. all particles have the same phase),
corresponding to the superfluid phase obtained for the unfrustrated system. In
the limit where correlated hopping dominates ($t' \simeq - 1$), the next-nearest 
neighbor antiferromagnetic interactions between the $xy$-components 
of the spins are minimized by having a N\'eel ordering of the $xy$-plane components of the spins 
in each square sublattice (SF 1). The
directions of the spins in the two sublattices are independent. This can be
understood by thinking of the simple hopping as a ferromagnetic nearest
neighbor interaction between the $xy$ components of the spins. For
each spin ${\bf S}_r$, the $xy$ components of the two spins
${\bf S}_{r + x}$ and ${\bf S}_{r + y}$ are in opposite directions.
Therefore $S^x_r S^x_{r + x} + S^y_r S^y_{r + x} = - (S^x_r S^x_{r + y} +
S^y_r S^y_{r + y})$ and the contributions in $x$ and $y$ directions simply
cancel each other. As the simple hopping is the only coupling in the $xy$-plane 
between the two sublattices, they become effectively decoupled.

When $V$ increases, a first supersolid phase is stabilized by the nearest
neighbor repulsion at densities $n>1/2$. This phase (SS 1) appears in the
region between the superfluid (SF 1) and the $n = 1$ insulating
phase and extends towards smaller values of the chemical potential as $V$
increases. As depicted in Fig.~\ref{fig:srcu2bo32:exactdiagonalization_negative_tp_phases}, the 
SS1 supersolid phase is based on a checkerboard solid order with one sublattice completely
filled and the other one which is partially filled having a N\'eel ordering
in the $xy$-plane. This phase clearly minimizes the
correlated hopping which acts as a next-nearest neighbor antiferromagnetic
interaction between the $xy$-plane components of the spins. At larger $V$,
another supersolid phase (SS 2) appears in the middle of the phase diagram at
densities $n < 1 / 2$. This phase is also based on a checkerboard solid order
but the $xy$-plane components of the spins are ferromagnetically ordered.
Between $V = 1$ and $V = 1.5$, a $n=1/2$ solid phase with checkerboard
order is stabilized in the middle of the phase diagram. As $V$ continues to
increase, the superfluid phases are slowly replaced by solid and supersolid
(SS 1) phases.

If one compares these results to those of the unfrustrated model ($t, t' > 0$),
one can observe several differences. As expected the superfluid regions are
smaller at intermediate values of the correlated hopping ($t' \simeq - 0.5$)
while the solid regions are larger and appear at a smaller value of the
nearest neighbor repulsion $V$. In contradiction to what was expected, the
supersolid phase (SS 1) is larger and can also be stabilized even when the
nearest neighbor repulsion is not large enough to stabilize the solid phase.
But this can be easily understood
by looking at the structure of the supersolid depicted in Fig.~\ref{fig:srcu2bo32:exactdiagonalization_negative_tp_phases}. Because one of
the sublattices is completely filled, the spins have no $xy$-plane components
in this sublattice and the simple hopping term has no effect on this state.
The correlated hopping term can therefore be fully minimized by having the
other sublattice with a N\'eel ordering in the $xy$-plane. In
conclusion, only the superfluid phases feel the frustration of the
system and become unstable, leaving a phase diagram dominated by supersolid
and solid phases. In comparison to the unfrustrated model ($t, t' > 0$), this
phase diagram is much more complex, with several different phases
characterized by interesting superfluid orders. This is clearly a consequence
of the frustration induced by the competition between simple hopping and
correlated hopping.

Regarding the validity of the results, the frustration might become
 a challenge for the CA. Indeed, it could stabilize other type of phases, which cannot be
described by the CA. It would be therefore interesting to
compare these results to the predictions of other methods. QMC
suffers from the sign problem, but one could use for example exact diagonalizations. 
This is left for future investigation however.

\section{Next-nearest neighbor hopping}

The previous section has shown that correlated hopping strongly favors supersolid phases. 
It would be interesting to see to what extend a standard hopping over the diagonal differs from a
 correlated hopping. We therefore study in the following the case $(t_2\neq 0;t'=0)$. Note that the model is now 
particle-hole symmetric as already mentioned above.

In the magnetic language used in the CA, the nearest neighbor hopping
becomes a ferromagnetic (antiferromagnetic) interaction between the next-nearest neighbor spins in the $xy$-plane 
when $t_2 > 0$ $(t_2 < 0)$. The energy scale will be fixed by $t + |t_2 | = 1$. Note that the unfrustrated (ferromagnetic) side
of this model has been studied by Chen and collaborators {\cite{chen08}} by QMC. This enables us to compare the numerically unbiased
results with the CA we are applying. Additionally, we will look again at the frustrated side as in the case
of correlated hopping studied in the last section.

\subsection{Unfrustrated Case $(t>0;t_2 >0)$}

\begin{figure}[t!]
  \begin{center}
    \includegraphics[width=0.9\textwidth]{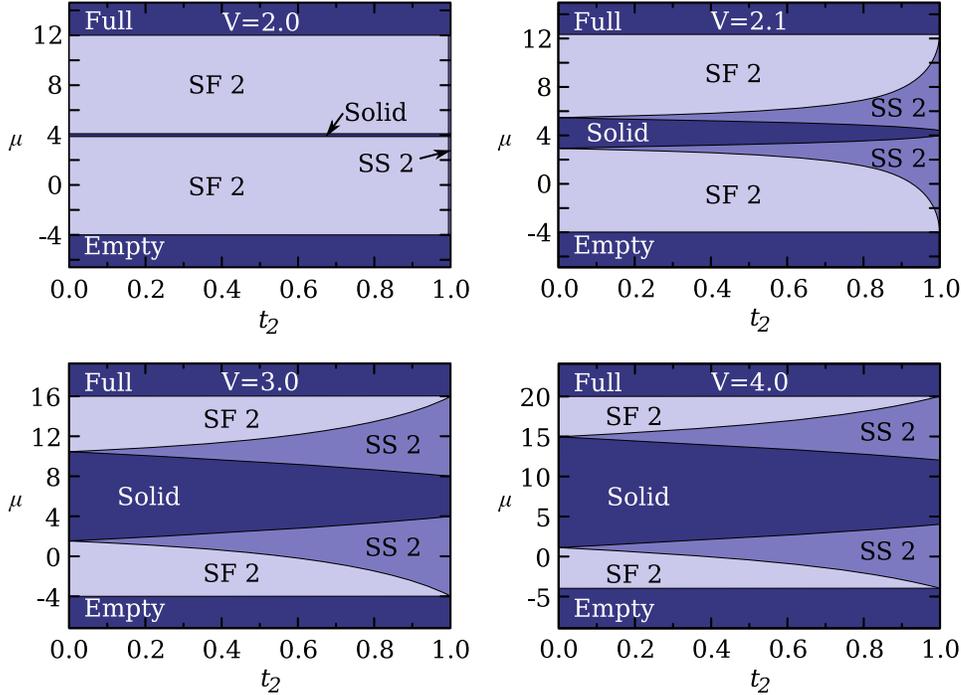}
  \end{center}
  \caption{\label{fig:srcu2bo32:exactdiagonalization_t2_phasediagram}Zero-temperature
  phase diagrams obtained with a classical approximation as a function of the
  next-nearest hopping $t_2 > 0$ and the chemical potential $\mu$ for various
  nearest neighbor repulsions $V$. All the transitions are second order, except
  at $t_2 = 0$ where the transition between superfluid and solid is first
  order.}
\end{figure}

\begin{figure}[t!]
  \begin{center}
    \includegraphics{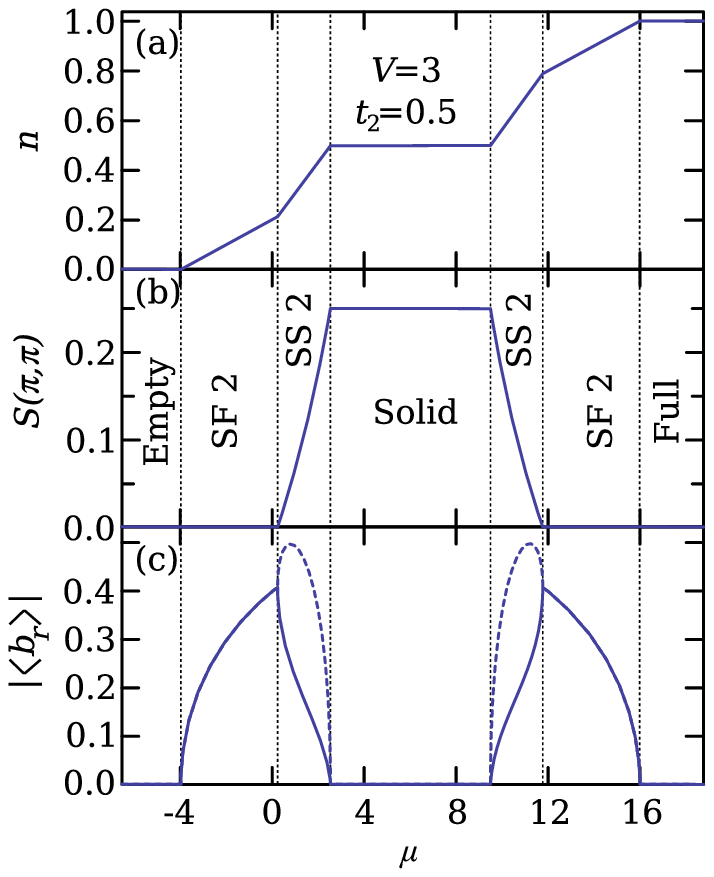}
  \end{center}
  \caption{\label{fig:srcu2bo32:exactdiagonalization_t2_orderparameters} Classical results for (a)
  the density $n$, (b) the static structure factor $S (\pi, \pi)$ and (c) the superfluid
  order parameter $| \langle b_r \rangle |$ for $r$ on the $A$ sublattice
  (solid lines) and $B$ sublattice (dashed lines) as a function of the
  chemical potential for $V = 3$ and $t_2 = 0.5$.}
\end{figure}

The zero-temperature classical phase diagram is shown in Fig.~\ref{fig:srcu2bo32:exactdiagonalization_t2_phasediagram} as a function of the
next-nearest neighbor hopping $t_2$ and the chemical potential $\mu$ for
various values of the nearest neighbor repulsion $V$. The density and the
order parameters are shown in Fig.~\ref{fig:srcu2bo32:exactdiagonalization_t2_orderparameters} as a function of
the chemical potential for $t_2 = 0.5$ and $V = 3$. When the nearest neighbor
repulsion is smaller than $2$ (not shown), the phase diagram displays only a
superfluid phase with a ferromagnetic ordering of the $xy$-plane components
of the spins, as depicted in Fig.~\ref{fig:srcu2bo32:exactdiagonalization_negative_tp_phases} (SF 2). When $V =
2$, a $n = 1 / 2$ checkerboard solid phase appears at $\mu = 4$ which is
stable for all values of the next-nearest neighbor hopping $t_2 \geqslant 0$.
Simultaneously, a supersolid phase appears at $t_2 = 1$. It is stable at
all densities except $n=1/2$. This phase has a checkerboard solid order
and the $xy$-plane components of the spins are ferromagnetically ordered, as
depicted in Fig.~\ref{fig:srcu2bo32:exactdiagonalization_negative_tp_phases}
(SS 2). When $V$ increases, the main effect is that the solid and supersolid
phases slowly replace the superfluid phase. Note that these results are consistent with the QMC results by Chen and
collaborators\cite{chen08}.

Comparing these results to those of the model with correlated hopping, one can
observe several similarities but also some interesting differences. The phases
stabilized by both models are the same, with a solid order which is always
based on a checkerboard pattern and a superfluid order which is always
uniform. This is not surprising as the solid order is a consequence of the
nearest neighbor repulsion which is the same for both models and the uniform
superfluid order is a consequence of the kinetic terms which, for both models,
act as ferromagnetic interactions in the $xy$-plane.
Another similarity is the strong tendency to stabilize large supersolid
regions when the kinetic energy is dominated by either correlated hopping or
next-nearest neighbor hopping once the solid phase is present. Finally, the phase transitions at densities $n>1/2$
 are second order in both models.

Regarding the differences between the two models, the supersolid phases
stabilized by correlated hopping appear only at densities $n>1/2$, while
they appear at any densities with next-nearest neighbor hopping. This can be
understood by realizing that the correlated hopping corresponds to a next-nearest 
neighbor hopping if a neighboring site is occupied, while it has no
effect if the neighboring sites are empty. Therefore at high densities in a
checkerboard supersolid correlated hopping and next-nearest neighbor hopping
have basically the same effect. This is no longer true for correlated 
hopping at densities $n<1/2$. In contrast, the model with normal diagonal hopping is
particle-hole symmetric and one obtains a supersolid by doping particles or holes.
This argument also explains why the phase transitions differ for
densities $n<1/2$. They are second order with a finite next-nearest
neighbor hopping and first order with correlated hopping. 

Finally, the most important difference is that correlated hopping can stabilize a supersolid
phase even when the nearest neighbor repulsion is too small to stabilize a
solid phase, while with next-nearest neighbor hopping the supersolid phases
are only stabilized with an adjacent solid phase\cite{schmi08}.

\subsection{Frustrated case $(t>0;t_2<0)$}

\begin{figure}[t!]
  \begin{center}
    \includegraphics[width=0.9\textwidth]{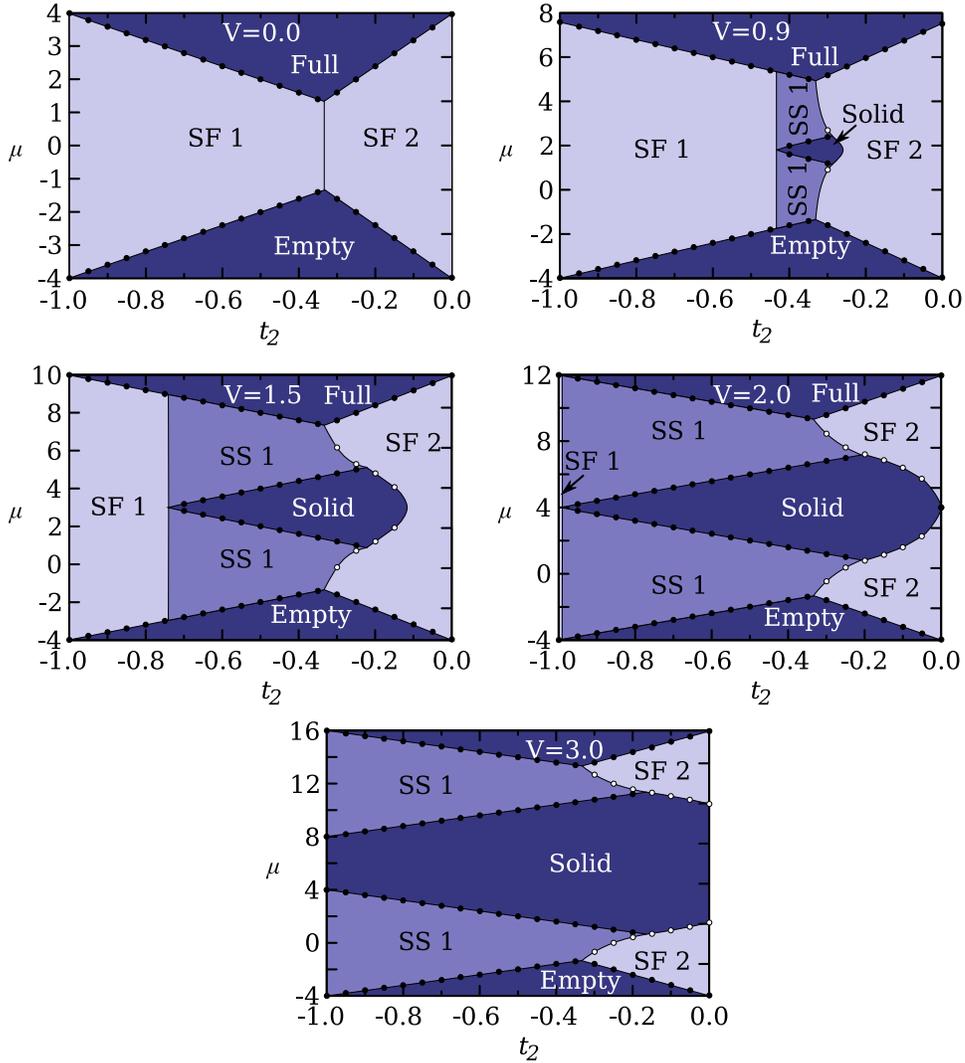}
  \end{center}
  \caption{\label{fig:srcu2bo32:exactdiagonalization_negative_t2_phasediagram}Zero-temperature
  phase diagrams as a function of the next-nearest neighbor hopping $t_2 < 0$
  and the chemical potential $\mu$ for various nearest neighbor repulsion $V$.
  White (black) circles correspond to first (second) order transitions
  obtained with the CA. The lines are guides to the
  eye. }
\end{figure}

\begin{figure}[t!]
  \begin{center}
    \includegraphics[width=0.9\textwidth]{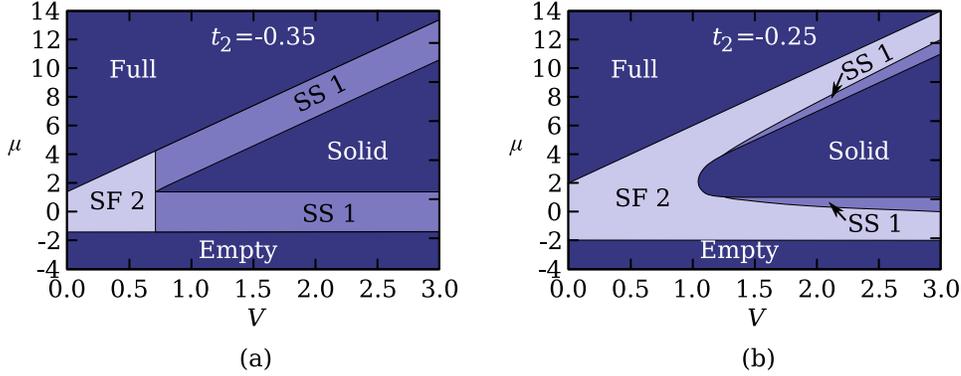}
  \end{center}
  \caption{\label{fig:srcu2bo32:exactdiagonalization_negative_t2_phasediagram_tp}Zero-temperature
  phase diagrams as a function of the nearest neighbor repulsion $V$ and the
  chemical potential $\mu$ at fixed next-nearest neighbor hoping $t_2 = -
  0.35$ (a) and $t_2 = - 0.25$ (b). }
\end{figure}

\begin{figure}[t!]
  \begin{center}
    \includegraphics[width=0.9\textwidth]{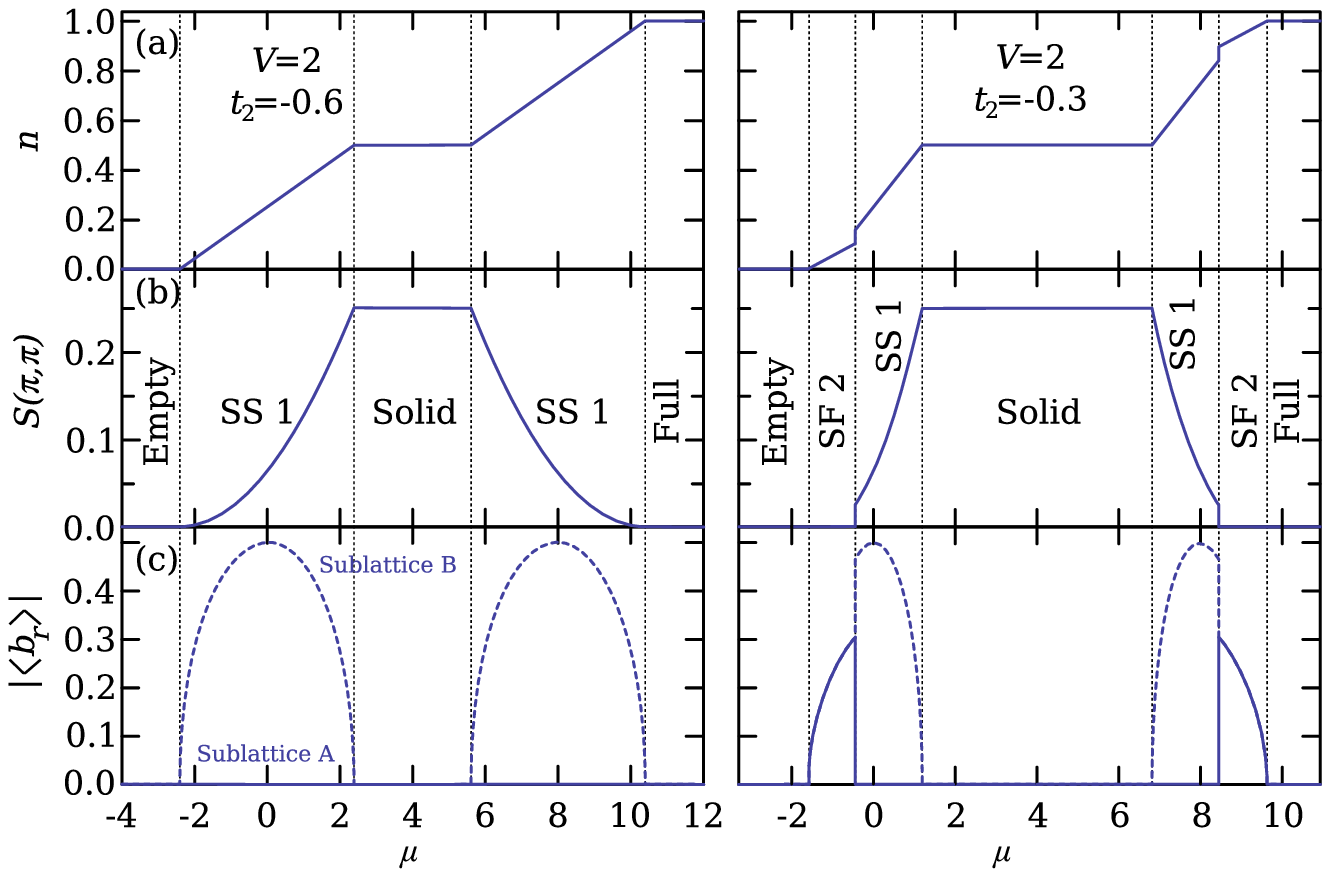}
  \end{center}
  \caption{\label{fig:srcu2bo32:exactdiagonalization_negative_t2_orderparameters}
  Classical results for (a)
  the density $n$, (b) the static structure factor $S (\pi, \pi)$ and (c) the superfluid
  order parameter $| \langle b_r \rangle |$  for $r$ on the $A$ sublattice
  (solid lines) and $B$ sublattice (dashed lines) as a function of the
  chemical potential for two representative cases. Left panel: $(t'=-0.6;V = 2)$. 
  Right panel: $(t'=-0.3;V = 2)$.}
\end{figure}

Fig.~\ref{fig:srcu2bo32:exactdiagonalization_negative_t2_phasediagram}
presents the classical phase diagram as a function of the next-nearest
neighbor hopping $t_2$ and the chemical potential $\mu$ for various values of
the nearest neighbor repulsion $V$. The density and the order parameters are
shown in Fig.~\ref{fig:srcu2bo32:exactdiagonalization_negative_t2_orderparameters} as a
function of the chemical potential $\mu$ for two representative cases. When $V = 0$, the
phase diagram is dominated by two superfluid phases. In the region where
nearest neighbor hopping dominates, the system stabilizes a superfluid phase
with a ferromagnetic ordering in the $xy$-plane,
corresponding to the superfluid phase obtained with $t_2 > 0$ which is depicted in 
Fig.~\ref{fig:srcu2bo32:exactdiagonalization_negative_tp_phases} (SF 2).
This phase clearly minimizes the nearest neighbor hopping which acts as a
ferromagnetic coupling between the $xy$-plane components of nearest neighbor
spins. In the other limit, where the next-nearest neighbor hopping dominates, the
system stabilizes a superfluid phase with a N\'eel ordering of the $xy$-plane
components of the spins in each sublattice, as depicted in Fig.~\ref{fig:srcu2bo32:exactdiagonalization_negative_tp_phases} (SF 1). 
In this phase the directions of the spins in the two sublattices are independent (as discussed before for the case of 
a frustrated correlated hopping).
This phase minimizes the next-nearest neighbor hopping which acts as an
antiferromagnetic coupling between the $xy$-plane components of the next-nearest
neighbor spins. When $V > 0.7$, a solid phase with checkerboard order appears
at half filling for intermediate values of $t_2$. Simultaneously, the
system realizes two supersolid phases, separated by the solid phase. The first
one, which appears at densities $n > 1 / 2$, has a checkerboard solid order
with one sublattice completely filled and the other which is partially filled
and presents a N\'eel ordering of the $xy$-plane components of the spins, as
depicted in Fig.~\ref{fig:srcu2bo32:exactdiagonalization_negative_tp_phases}
(SS 1). The second supersolid phase has also a checkerboard solid order but
one sublattice is completely empty and the other is partially filled with a
N\'eel ordering in the $xy$-plane. Its presence is a direct consequence of the 
particle-hole symmetry of the model. The size of the solid
and supersolid regions increase upon increasing $V$, and they slowly replace the superfluid
phases. In order to complement the understanding of the whole phase diagram, Fig.~\ref{fig:srcu2bo32:exactdiagonalization_negative_t2_phasediagram_tp}
presents the zero-temperature phase diagram as a function of the nearest
neighbor repulsion $V$ and the chemical potential $\mu$ for a next-nearest
neighbor hopping $t_2 = - 0.35$ (a) and $t_2 = - 0.25$ (b). Note that the
information is the same as in Fig.~\ref{fig:srcu2bo32:exactdiagonalization_negative_t2_phasediagram}.

A comparison to the model with correlated hopping leads to essentially the
same conclusions as for the unfrustrated case. Both
systems stabilize the same type of phases and have a strong tendency to
stabilize supersolid phases when the kinetic energy is dominated by $t_2$ or
$t'$. But again, the normal hopping over the diagonal does not stabilize a supersolid phase
without an adjacent solid phase.

\section{Conclusion}

In this work we studied a $t-V$-model of interacting hard-core bosons on the square lattice
with either a correlated or a standard next-nearest neighbor hopping over the diagonal. 
We studied the full phase diagram within a classical approximation, focusing on
the realization of supersolid phases. 

Both correlated hopping and standard diagonal hopping 
favor large supersolid phases once a checkerboard solid is realized. Here it does
not matter whether the different kinetic processes are frustrated or unfrustrated. 
In contrast when the solid phase is {\it not} stabilized, only the correlated hopping allows to have 
a supersolid phase without any adjacent solid phase present\cite{schmi08}. 
All these results obtained within the classical approximation are in qualitative agreement
with QMC simulations\cite{schmi08,chen08}.

We also studied the case of frustrated kinetic couplings which is not accessible by
QMC simulations. For both type of kinetic processes, 
the phase diagram is more complex than in the unfrustrated case. It shows various competing 
superfluid and supersolid phases. The superfluid phases become less stable because of the
frustration which leads to a phase diagram which is mostly dominated by solid and
supersolid phases. 

In conclusion, the phase diagram of hard core bosons on the square
lattice revealed by a classical approximation is extremely rich, with several solid and
supersolid phases, as soon as correlated hopping or next-nearest neighbor hopping
is included. We hope that the present results will stimulate the experimental investigation
of frustrated dimer-based quantum magnets in a magnetic field since they naturally leads
to such additional kinetic processes. 

\section*{Acknowledgements}
We acknowledge very useful discussions with A.~Laeuchli. KPS acknowledges ESF and EuroHorcs for funding through his EURYI. Numerical simulations were done on Greedy at EPFL ({\it greedy.epfl.ch}). This work has been supported by the SNF and by MaNEP.

\end{document}